\documentclass[aps,prd,twocolumn,10pt,aps,amsmath,amssymb,nofootinbib,superscriptaddress]{revtex4-2}
\usepackage{epsfig} 
\usepackage{amsmath} 
\usepackage{graphicx}
\usepackage{color}
\usepackage{float}
\usepackage{soul,xcolor}
\usepackage[font=scriptsize]{caption}
\usepackage{braket}
\usepackage{bbold}
\usepackage{subfig}
\usepackage{braket}
\usepackage{titlesec}
\usepackage{placeins}
\usepackage{hyperref}
\usepackage{caption}
\usepackage[bottom]{footmisc}
\begin{document}
\title{ Impact of Scalar NSI on Spatial and Temporal Correlations in Neutrino Oscillations
}

 \author{Bhavna Yadav}
\email{yadav.18@alumni.iitj.ac.in} 
\affiliation{Indian Institute of Technology Jodhpur, Jodhpur 342037, India}
\affiliation{Wilczek Quantum Center, Shanghai Institute for Advanced Studies, University of Science and Technology of China, Shanghai 201315, China}

\author{Ashutosh Kumar Alok}
\email{Deceased }
\affiliation{Indian Institute of Technology Jodhpur, Jodhpur 342037, India}

\begin{abstract}

Neutrino oscillation experiments are gradually approaching an era of precision, where subleading effects can also be tested. One such subleading effect is Non-Standard Interactions (NSI), which can play a crucial role in neutrino oscillations. Various works have typically discussed vector NSI in the context of quantum correlations. Recently, there have been improvements in the bounds on scalar NSI as well. In light of these developments, we aim to examine the impact of scalar NSI on quantum correlation measures. To analyze this impact, we are considering the strongest measure of quantum correlation, i.e., non-locality. Our study will encompass both spatial and temporal non-locality measures. This work presents the first investigation of scalar NSI in the context of quantum correlation measures.
\end{abstract}

\maketitle

\section{Introduction} \label{s:intro}

The non-classical nature of quantum systems has been a focal point of investigation since the inception of quantum physics. Schrödinger first introduced the notion of quantum correlation in 1935 \cite{schrodinger_1935}. Among the intriguing features of the quantum domain is nonlocality, famously derided by Einstein as ``spooky actions at a distance." This concept was initially presented in the 1935 EPR paradox by Einstein, Podolsky, and Rosen \cite{Einstein:1935rr}. John Bell's 1964 theorem \cite{Bell:1964kc} provided a theoretical foundation for nonlocality, later corroborated by experiments conducted by Clauser and Freedman in 1972, and Aspect in 1982 \cite{aspect}. John Bell analytically examined the compatibility of Einstein's theory of local hidden variables with the principles of locality and realism, establishing the Bell-CHSH (Clauser-Horne-Shimony-Holt) inequality as a critical test \cite{Bell:1964kc}. He showed that the correlations between measurement results from two spatially separated systems must adhere to a specific inequality, now known as the Bell-CHSH inequality \cite{chsh}. The violation of this inequality indicates the presence of nonlocal correlations, thus revealing their quantum nature. Nonlocality describes the phenomenon where particles can instantaneously affect each other's states regardless of the distance separating them.  Quantum correlations, grounded in the principles of locality and realism, provide a framework for testing local hidden variable theories and serve as effective tools for distinguishing classical from quantum behavior.

In 1985 a set of new inequalities known as Leggett-Garg inequalities (LGIs) were developed \cite{Leggett}. These inequalities are based on the assumptions of macrorealism (MR) and noninvasive measurement (NIM), positing that any system demonstrating macroscopic realism must comply with them. LGIs are the temporal counterparts of Bell's inequalities, examining correlations within a single system measured at different times, thus enabling a thorough examination of quantum mechanics on a macroscopic scale. The Leggett-Garg Temporal Inequality (LGtI), a variation of LGIs, is formulated by replacing the NIM criterion with a less stringent condition known as stationarity \cite{Emary}. Violations of the LGtI indicate a departure from macroscopic realism, revealing non-classical temporal correlations that are inherently quantum in nature. Such violations suggest that the behavior of the system cannot be mimicked by any theory that relies on the assumptions of macrorealism.

Most of those measures are studied in the context of neutrino oscillation \cite{PRD77, blasone2009al, Blasone2014, ALOK2016, banerjee2015quantum,banerjee2016quantum, Formaggio:2016cuh, Naikoo:2017fos, Naikoo:2019eec, Fu:2017hky,roy, Hu,Ding, kdcoherence,Ming:2020nyc,Ming2021,shang,xie, Sarkar:2020vob,pm, yadav,single,Jha,jha1, wang, ch2023am, ettefaghi, dixit3, soni, Konwar:2024nrd, Konwar:2024nwc}, most of them consider vacuum, and some of them also include matter effects in Standard Model (SM). There are few works that study NSI. However, it is restricted to vector NSI.

 Neutrino oscillation is a well-known phenomenon supported by multiple experimental results \cite{fukuda1998, abe2016, aharmim2013, araki2005}. Neutrino oscillation is significant because of its implications in determining the mass of neutrinos, which in turn could lead to a revision of the SM of particle physics. Neutrino oscillations can only occur if neutrinos have a mass \cite{camilleri2008}. Although the three-flavor neutrino oscillation framework successfully explains the majority of the existing oscillation data, various novel physics scenarios may emerge as a result of BSM physics that may significantly affect neutrino oscillations \cite{argue1,argue2}. One such new physics scenario could result in a non-standard interaction (NSI) of neutrinos with ordinary matter \cite{Wolfenstein, babu, zhou}, and it is one of those new physics scenarios that might significantly change neutrino oscillations. As a result, it is essential to look into the consequences of NSI. The NSIs are classified into two types: vector NSI and scalar NSI. In the case of vector NSI, vector bosons mediate new interactions and allow for parameterization with vector current, similar to the ordinary matter effect but with unknown couplings \cite {Wolfenstein}, whereas in the case of scalar NSI, neutrinos can couple to the scalar fields and cause a correction to the neutrino mass term \cite {parke}. There is currently a rising interest in investigating the impacts of scalar NSI in experiments studying neutrino oscillations \cite{parke, xu, khan, moon,denton, medhi,gupta, essnub, mohanta, DENTON2024, Sarker1, Sarker2}. 

Efforts are being made to elucidate the solar neutrino data obtained from the Borexino experiment \cite{borexino} using the oscillation framework that incorporates scalar NSI \cite {parke}. Constraints on off-diagonal scalar NSI parameters are obtained from the data of T2K and NO$\nu$A experiments \cite{denton}. Big Bang nucleosynthesis provides some constraints on scalar NSI parameters \cite{venzor}, and different astrophysical and cosmological constraints have also been used in studies \cite{babu2}. In this work, we consider an extensive study of the effects of scalar NSI on several measures of quantum correlations. This includes measures that capture nonlocality.

The layout of the paper is as follows: In Sec. II, we provide a very brief review of measures of quantum correlations. Sec. III outlines the formalism of neutrino oscillation in SM with vector and scalar NSI. In Sec. IV, we present our results and the conclusion provided in Sec. V.

\section{Measures of Non Locality }
This section provides a concise overview of the non-locality measures implemented in the present analysis. We will mainly focus on CHSH inequality and LGtI.
\\
\\
\textbf{\emph{ CHSH (Clauser-Horne-Shimony-Holt) Inequality:} } A Bell-type inequality, the most well-known of which is the CHSH inequality, can be violated to disclose the spatial quantum nonlocality of two-qubit systems.  The CHSH inequality is expressed as follows \cite{wang, qin}:
\begin{equation}
    B=\left| \left<\mathcal{B} \right>_{\rho }\right|=\left|Tr\left ( \rho \mathcal{B} \right ) \right|\leq 2,
\end{equation}
where $\mathcal{B}$ is Bell operator corresponding to CHSH inequality, can be written as
\begin{equation}
   \mathcal{B}=A_{1}\otimes B_{1}+A_{1}\otimes B_{2}+A_{2}\otimes B_{1}-A_{2}\otimes B_{2}, 
\end{equation}
where $A_{i}=\vec{a}_{i}.\vec{\sigma_{A} }$, $B_{j}=\vec{b}_{j}.\vec{\sigma_{B} }$, and $\sigma's$ are Pauli matrices, $\vec{a}_{i}$ and $\vec{b}_{j}$ are real unit vectors. For two-qubit density matrix $\rho$, $\left<CHSH\right>_{\rho }$ i.e. maximal mean value of $\left<\mathcal{B}\right>_{\rho }$
under all possible measurements, defined as
\begin{equation}
    \left<CHSH\right>_{\rho }=max_{A_{i},B_{j}}Tr\left ( \rho \mathcal{B} \right )={\color{black}2} \sqrt{n_{1}+n_{2}},
\end{equation}
where $n_{1}$ and $n_{2}$ are two largest eigenvalues of matrix $M^{\dagger}M$ and M is correlation matrix; $m_{ij}=Tr\left ( \rho \sigma_{i}\otimes \sigma_{j} \right)$ are the elements of this correlation matrix. By using pairwise bipartite states, a trade-off relation can be obtained for three-qubit state $\rho_{ABC}$,
\begin{equation}
    \left< CHSH\right>_{\rho_{AB}}^{2}+\left<CHSH\right>_{\rho_{AC}}^{2}+\left<CHSH\right>_{\rho_{BC}}^{2}\leq 12,
\end{equation}
where $\rho_{AB}$, $\rho_{AC}$ and $\rho_{BC}$ are reduced density matrices. This relation can be used to study the nonlocality in a three-qubit system. For the flavour neutrino system, these squares of CHSH of the reduced density matrix can be written in terms of probabilities as
\begin{eqnarray}
    \left< CHSH\right>_{\rho_{AB}^{\alpha }}^{2}={\color{black}4}\left\{ 4 P_{\alpha e}P_{\alpha \mu}+max\left [ 4 P_{\alpha e}P_{\alpha \mu},\left ( 2 P_{\alpha \tau}-1 \right )^{2} \right ]\right\},\notag\\
    \left< CHSH\right>_{\rho_{AC}^{\alpha }}^{2}={\color{black}4}\left\{ 4 P_{\alpha e}P_{\alpha \tau}+max\left [ 4 P_{\alpha e}P_{\alpha \tau},\left ( 2 P_{\alpha \mu}-1 \right )^{2} \right ]\right\},\notag\\
    \left< CHSH\right>_{\rho_{BC}^{\alpha }}^{2}={\color{black}4}\left\{ 4 P_{\alpha \mu}P_{\alpha \tau}+max\left [ 4 P_{\alpha \mu}P_{\alpha \tau},\left ( 2 P_{\alpha e}-1 \right )^{2} \right ]\right\}.
\end{eqnarray}
where $\alpha$ is the initial flavour state of neutrino. The trade-off relation states that the total maximal violation of the CHSH inequality tests for three qubits of the reduced bipartite neutrino flavor states must be less than or equal to 12.
\\
\textbf{\emph{Leggett-Garg type Inequality (LGtI):} } The Leggett-Garg inequalities, grounded in the assumptions of macro-realism and noninvasive measurement, aim to elucidate correlations between measurements conducted on a system at different points in time. Macro-realism posits that a macroscopic system with multiple distinct states invariably exists in one of those states, whereas noninvasive measurement asserts the feasibility of performing measurements on a system without perturbing its dynamics. LGIs were originally introduced to elucidate macroscopic coherence by investigating the application of quantum mechanics to a many-particle system undergoing decoherence \cite{Leggett}. Additionally, LGI tests offer a means to scrutinize the concept of realism, incorporating hidden variable theories that propose predetermined values for a system's parameters regardless of measurement \cite{Huelga, Emary}. The violation of these inequalities signifies that such hidden variable theories cannot serve as alternatives for describing the temporal evolution of a quantum mechanical system. A linear combination of autocorrelation functions can be used to express the LGI parameter $K_{n}$ as \cite{Naikoo:2017fos}: \begin{equation}
    K_n = \sum_{i=1,j=i+1}^{i=n-1}C(t_i,t_j)- C(t_1,n),
\end{equation}
with $C(t_i,t_j)=\langle\hat{Q}(t_i)\hat{Q}(t_j)\rangle $  $=$ $Tr[\{\hat{Q}(t_i),\hat{Q}(t_j)\}\rho(t_0)$],
Here the average is taken with respect to $\rho(t_0)$, initial state of the system at time t=0, $\hat{Q}$ is a generic dichotomic operator, i.e. $\hat{Q}=\pm 1$, $\hat{Q}^+= Q$ and $\hat{Q}^2=1$ with $\hat{Q}= 1$ if the system is in the target state; otherwise, $\hat{Q}=- 1$. For $n\geq 3$ if realism is satisfied, then LGI parameter $K_n\leq n-2$. For n=3, LG inequality can be written as \begin{equation}
    K_{3}= C(t_{1}, t_{2})+C(t_{2},t_{3})-C(t_{1},t_{3}) \leq 1.
\end{equation}
Given that the correlation function $C(t_{2},t_{3})$ is contingent upon a comprehensive measurement of the system, this dependency contradicts the NIM postulate. To address this issue, the NIM postulate has been replaced with a less stringent condition of stationarity. By applying this weakened condition, the correlation function $C(t_{i},t_{j})$ becomes contingent solely on the time difference. {\color{black}We choose the operator $\hat{Q}$ that measures whether the neutrino is in a specific flavor state 
$\ket{\nu_\beta}$ ($\beta = e, \mu, \tau$) or not:
\begin{equation}
    \hat{Q} = 2\,\ket{\nu_\beta}\bra{\nu_\beta} - I.
\end{equation}
This operator has eigenvalues $Q = +1$ (if the neutrino is found in flavor $\beta$) and 
$Q = -1$ (otherwise). Under the assumption of stationarity, the two-time correlation function 
$C(t_i, t_j)$ depends only on the time difference.} Assuming a time interval such that $t_{2} - t_{1} = t_{3} - t_{2} = t$ and setting $t_{1}=0$, the LGtI parameter $(K_{3})$ can be expressed as follows \cite{Naikoo:2017fos, Formaggio:2016cuh, Naikoo:2019eec} :
\begin{equation}\label{lg}
    K_3= 2C(0,t)-C(0,2t)\leq 1.
\end{equation}
{\color{black}For a neutrino initially produced in the flavor state 
$\ket{\nu_\alpha}$, this correlation function can be compactly expressed in terms of the neutrino oscillation probability 
$P_{\alpha\beta}(t) = |\braket{\nu_\beta | \nu_\alpha(t)}|^2$ as follows~\cite{Naikoo:2019eec}:
\begin{equation}
    C(0, t) = 1 - 2 P_{\alpha\beta}(t).
\end{equation}
Substituting this expression into Eq.~\ref{lg} for $n = 3$ immediately yields the LGtI parameter $K_3$ 
in terms of the oscillation probability, as }
\begin{equation}
    K_3= 1+2P_{\alpha,\beta}(2t,E)- 4P_{\alpha,\beta}(t,E).
\end{equation}
In the ultrarelativistic limit t=L the LGtI  parameter $K_3$ is given as:
\begin{equation}
   K_3= 1+2P_{\alpha,\beta}(2L,E)- 4P_{\alpha,\beta}(L,E),
\end{equation}
which shows experimental feasibility of LG function on using condition $P_{\alpha, \beta}(2L,E)$ = $P_{\alpha, \beta}(L,\Tilde{E})$ with suitable choice of E and $\Tilde{E}$ \cite{Formaggio:2016cuh}. {\color{black} The condition $P_{\alpha\beta}(2L, E) = P_{\alpha\beta}(L, \bar{E})$ is an approximation that enables the experimental test of the LGtI. Directly measuring the neutrino state at three successive times (or baselines $0, L, 2L$) is not feasible, as detection is typically destructive. Instead, the energy dependence of the oscillation probability is used as a proxy for its baseline dependence. The relationship is motivated by the form of the oscillation phase, $\phi_{ij} \propto \frac{\Delta m_{ij}^2 L}{E}$. A ``suitable choice'' of energy $\bar{E} \approx E$ at a fixed baseline $L$ provides a probability $P_{\alpha\beta}(L, \bar{E})$ that approximates the probability $P_{\alpha\beta}(2L, E)$ that would be required for a direct test. This allows for a practical test of the LGtI by measuring probabilities at different energies and a single baseline.}

\section{Formalism} \label{sec2}


In the standard formalism of three-flavor neutrino oscillations, the flavor states $\nu_{e}$, $\nu_{\mu}$ and $\nu_{\tau}$ are represented as a linear superposition of the mass eigenstates $\nu_{1}$, $\nu_{2}$ and $\nu_{3}$:
{\color{black}\begin{equation}\label{1}
    \ket{\nu_{\alpha}}= \sum_{i}U_{\alpha i}^*\ket{\nu _{i}}, 
\end{equation}}
where $\alpha = e, \mu , \tau $ and $i = 1, 2 , 3 $. The flavour state $\ket{\nu_{\alpha}}$ denotes the initial flavour state at $t= 0$, which due to mixing, can be expressed in terms of mass eigenstates $\ket{\nu_{i}}$. The $3\times 3$ mixing matrix, known as Pontecorvo-Maki-Nakagawa-Sakata (PMNS)matrix, is characterized by three mixing angles i.e., $\theta_{12}$, $\theta_{13}$, $\theta_{23}$ and a CP (charge-parity) phase. The matrix elements of the $3\times 3$ PMNS mixing matrix can be defined as
\begin{equation}\label{2}
\begin{pmatrix}
c_{12}c_{13} & s_{12}c_{13}&s_{13}e^{-\iota\delta}\\ 
-s_{12}c_{23}-c_{12}s_{13}s_{23}e^{\iota\delta}& c_{12}c_{23}-s_{12}s_{13}s_{23}e^{\iota\delta }&c_{13}s_{23}\\ 
s_{12}s_{23}-c_{12}s_{13}c_{23}e^{\iota\delta }& -c_{12}s_{23}-s_{12}s_{13}c_{23}e^{\iota\delta}&c_{13}c_{23}
\end{pmatrix}, 
\end{equation}
where $s_{ij} = \sin \theta _{ij}$ and $c_{ij} = \cos \theta _{ij}$.  Mass eigenstate evolve as
\begin{equation}\label{3a}
     \ket{\nu_{i}(t)}=e^{-\iota  E_{i}t}\ket{\nu_{i }},
\end{equation}
where $\ket{\nu_{i}}$ are mass eigenstate at $t=0$ given by Eq. \eqref{3a}. Therefore, the time evolution of the flavour eigenstate  can be expressed as follows
{\color{black}\begin{eqnarray}\label{3}
    \ket{\nu _{\alpha}(t)}=\sum_{i}  U_{\alpha i}^* e^{-\iota  E_{i}t} \ket{\nu_{i}}.
\end{eqnarray}}
In the relativistic limit,  neutrino flavor states are treated as separate modes. A mode is a way to partition the quantum field into distinct segments, and each mode can encompass various states. When two modes become entangled, it suggests that their quantum states are correlated in some manner. Neutrino mode entanglement refers to a specific type of entanglement occurring between different modes within a neutrino field. Consequently, the time-evolved flavor state can be understood as an entangled combination of flavor modes. In the context of a three-flavor neutrino system, this concept can be represented as \cite{blasone2009al}
\begin{eqnarray}\label{m4}
    \ket{\nu _{e}}\equiv \ket{1}_{e} \ket{0}_{\mu }\ket{0}_{\tau }\equiv\ket{100}_{e \mu \tau}\notag \\
    \ket{\nu_{\mu }}\equiv\ket{0}_{e } \ket{1}_{\mu } \ket{0}_{\tau }\equiv\ket{010}_{e \mu \tau}\\
   \ket{\nu_{\tau }}\equiv\ket{0}_{e } \ket{0}_{\mu } \ket{1}_{\tau }\equiv\ket{001}_{e \mu \tau}\notag
\end{eqnarray}\\

Using Eq. (\eqref{3}) and \eqref{m4}, we get the time evolution for the tripartite quantum state as
\begin{eqnarray}\label{5}
    \ket{\nu _{\alpha }(t)}=\Bar{U}_{\alpha e}(t)\ket{100}_{e \mu \tau}+\Bar{U}_{\alpha \mu}(t) \ket{010}_{e \mu \tau}
    +\Bar{U}_{\alpha \tau}(t) \ket{001}_{e \mu \tau},
\end{eqnarray}
where $\alpha$ is the flavour state at the time $t=0$ and $\Bar{U}=U e^{-i \mathcal{H}_m t} U^{\dagger}$, $\mathcal{H}_m = diag(E_1,E_2,E_3)$ {\it i.e.,} the Hamiltonian in the mass eigenbasis with energies $E_{i}$ $(i=1,2,3)$. Eq.\eqref{5} delineates the entanglement at time $t$ among the flavor modes. Consequently, we can employ the framework of quantum resources to analyze the inherent quantum properties within the oscillating neutrino system.
\\
The density matrix corresponding to the state expresssed in Eq. \eqref{5}, denoted as $\rho _{ABC}^{\alpha}(t)= \ket{\nu _{\alpha }(t)}\bra{\nu _{\alpha }(t)}$, is formulated as
\begin{equation}{\label{rho1}}
    \rho _{ABC}^{\alpha}(t)=\begin{pmatrix}
0 & 0 & 0 & 0 & 0 & 0 & 0 & 0\\ 
0 & \rho _{22}^{\alpha } & \rho _{23}^{\alpha } & 0 & \rho _{25}^{\alpha } & 0 & 0 & 0\\ 
0 & \rho _{32}^{\alpha } & \rho _{33}^{\alpha } & 0 & \rho _{35}^{\alpha } & 0 & 0 & 0\\ 
0 & 0 & 0 & 0 & 0 & 0 & 0 & 0\\ 
0 & \rho _{52}^{\alpha } & \rho _{53}^{\alpha } & 0 & \rho _{55}^{\alpha } & 0 & 0 & 0\\ 
0 & 0 & 0 & 0 & 0 &0  & 0 & 0\\ 
0 & 0 & 0 & 0 & 0 &0  & 0 & 0\\
0 & 0 & 0 & 0 & 0 &0  & 0 & 0 
\end{pmatrix},
\end{equation}
where the elements of this matrix can be written as 
\begin{eqnarray}\label{rhop}
    \rho _{22}^{\alpha}= \left |\Bar{U}_{\alpha \tau }(t) \right |^{2};\hspace{0.3 cm}\rho _{23}^{\alpha}= \Bar{U}_{\alpha \tau }(t)\Bar{U}_{\alpha \mu }^{\ast }(t);\hspace{0.3 cm}\rho _{25}^{\alpha}= \Bar{U}_{\alpha \tau }(t)\Bar{U}_{\alpha e }^{\ast }(t);\notag \\
    \rho _{32}^{\alpha}= \Bar{U}_{\alpha \mu }(t)\Bar{U}_{\alpha \tau }^{\ast }(t);\hspace{0.3 cm}\rho _{33}^{\alpha}= \left |\Bar{U}_{\alpha \mu }(t) \right |^{2};\hspace{0.3 cm}\rho _{35}^{\alpha}= \Bar{U}_{\alpha \mu }(t)\Bar{U}_{\alpha e }^{\ast }(t); \\
     \rho _{52}^{\alpha}= \Bar{U}_{\alpha e }(t)\Bar{U}_{\alpha \tau }^{\ast }(t);\hspace{0.3 cm}\rho _{53}^{\alpha}= \Bar{U}_{\alpha e }(t)\Bar{U}_{\alpha \mu }^{\ast }(t);\hspace{0.3 cm}\rho _{55}^{\alpha}= \left |\Bar{U}_{\alpha e }(t) \right |^{2}.\notag
\end{eqnarray}
The corresponding probabilities are: $P_{\alpha e}(t)= \left |\Bar{U}_{\alpha e  }(t) \right |^{2}$, $P_{\alpha \mu}(t)= \left |\Bar{U}_{\alpha \mu  }(t) \right |^{2}$ and $P_{\alpha \tau}(t)= \left |\Bar{U}_{\alpha \tau  }(t) \right |^{2}$.

In the ultra-relativistic limit, length $L\equiv t$ and $E_i -E _j \approx \Delta m_{ij}^{2}/2 E$. 

When neutrinos travel through matter, they can interact via charged currents (CC) and neutral currents (NC), affecting the flavor oscillation pattern. Within a medium, considering these interactions, the Standard Model Hamiltonian can be formulated as \cite{Giunti:2007ry}
\begin{eqnarray}\label{osc3}
\mathcal{H}_m=\mathcal{H}_{vac}+\mathcal{H}_{mat}
=\begin{pmatrix}
E_{1} & 0 & 0 \\ 
0 & E_{2} & 0 \\
0 & 0 & E_{3}
\end{pmatrix}+
U^\dagger\begin{pmatrix}
A & 0 & 0 \\ 
0 & 0 & 0 \\
0 & 0 & 0 
\end{pmatrix}U,
\end{eqnarray}
 where $\mathcal{H}_{vac}$ and $\mathcal{H}_{mat}$ are Hamiltonian for vacuum and matter neutrino oscillations, respectively. Further $A=\pm \sqrt{2}G_{F}N_{e}$ is the matter potential. $G_{F}$ is the Fermi constant and $N_e$ is number density of electron in matter. The sign of matter potential depends upon the type of neutrinos. It is positive (negative) for neutrinos (antineutrinos).  {\color{black} In our analysis, we adopt the standard constant density approximation for the neutrino baseline. For the DUNE experiment, which is a long-baseline experiment where the neutrino beam propagates through the Earth's crust, we use a characteristic crust density of $\rho \approx 2.8~\text{g/cm}^{3}$, yielding a matter potential of $A \approx 10^{-13}~\text{eV}$.

Although matter effects can lead to the Mikheyev-Smirnov-Wolfenstein (MSW) resonance, it is important to note that this resonance does not occur within the DUNE energy range for the given crust density. The resonance energies for oscillations in matter can be estimated as~\cite{msw1}
\begin{equation}
    E_{\text{res}}(\text{GeV}) \approx 6.55 \times 10^{3} \,
    \frac{\Delta m_{ij}^{2}~(\text{eV}^{2}) \cos 2\theta_{ij}}{\rho Y_{e}},
\end{equation}
where $\rho$ is in g/cm$^{3}$ and $Y_{e}$ is the electron fraction. For the Earth's crust with $\rho = 2.8~\text{g/cm}^{3}$ and $Y_{e} \approx 0.5$, the resonance between $\nu_{1}$ and $\nu_{2}$ occurs at $E_{\text{res}}^{12} \approx 0.14~\text{GeV}$, and between $\nu_{1}$ and $\nu_{3}$ at $E_{\text{res}}^{13} \approx 11.17~\text{GeV}$. Since the DUNE energy spectrum is focused in the 1-10 GeV range, it lies between these two resonances, and thus significant MSW effects are not expected for this setup.
}

The time evolution operator describing the transition between neutrino mass eigenstates is encapsulated by $U_{m}(L)$. For three-flavor neutrino oscillations, the evolution operator $U_{m}(L)$ can be defined as \cite{Ohlsson:1999xb}  
\begin{eqnarray}\label{osc4}
U_{m}(L)=e^{-i\mathcal{H}_{m}L}=\phi ~e^{-iLT}~~~~~~~~~~~~~~~~~~~~~~~~~~~~~~~~~~\\=\phi \sum_{a=1}^{3}e^{-i L\lambda _{a}}\frac{1}{3\lambda _{a}^2+c_{1}}\left [( \lambda _{a}^{2}+c_{1}) I+\lambda _{a}T+T^{2}\right ].\notag
\end{eqnarray}
Here, \( \phi \) is a complex phase factor defined as  $\phi = e^{-iL (\operatorname{tr} H_{\text{m}})/3}$
where \( L \) represents the distance between the source and the detector, and \( H_{\text{m}} \) denotes the Hamiltonian, incorporating vacuum and matter effects. The coefficient \( c_1 \) is a real quantity given by the product of the determinant of \( T \) and the trace of \( T^{-1} \). The eigenvalues of the matrix \( T \) are represented by \( \lambda_1, \lambda_2, \) and \( \lambda_3 \), which are expressed as:  

\begin{equation}\label{osc5}
T \equiv \mathcal{H}_{vmac}- (tr H_{m})I/3 =
\begin{pmatrix}
T_{11} & T_{12}& T_{13} \\
T_{21} & T_{22}& T_{23} \\
T_{31} & T_{32}& T_{33} 
\end{pmatrix}.
\end{equation}

To determining neutrino oscillation probabilities, we employ the relation $U_{f}(L)=U^{\dagger} U_{m}(L) U$, where, $U_{f}(L)$ is an evolution operator in flavor basis. This operator facilitates the determination of oscillation probabilities between different neutrino flavors. To extract the relevant matrix element of the evolution operator in the flavor basis, we compute the absolute values of these elements and square them. This yields the oscillation probabilities for the transition from the initial flavor $\nu_{\alpha}$ to the final flavor $\nu_{\beta}$, which are represented as
\begin{equation}
   P_{\alpha \beta }\equiv \left|A_{\alpha \beta } \right|^{2}= \left| \bra\beta U_{f}(L) \ket\alpha\right|^{2}.
\end{equation}
\\
In the described formalism, the symbol $P_{\alpha \beta } $ represents the probability in matter. 
\\
\subsection{Vector NSI}
The effect of physics beyond the SM, notably NSI, emerges as a subleading factor in neutrino flavor oscillations \cite{Wolfenstein, Mikheev}. NSI effects, observable in both CC and NC interactions, are effectively described by four-fermion dimension-6 operators \cite{Farzan:2017xzy, Davidson:2003ha, Antusch:2008tz, Dev:2019anc,ohlsson13, Babu:2019mfe}. While the constraints on CC-NSI are stringent, necessitating their exclusion from our analysis, NC-NSI remains comparatively less restricted \cite{Biggio:2009nt}. NC-NSI effects can be parameterized as follows
\begin{eqnarray}\label{nsi1}
 \mathcal{L}_{NSI}^{NC}=2\sqrt{2}G_F\sum\limits_{\alpha, \beta, P} \epsilon_{\alpha\beta}^{f,P}(\bar{\nu}_\alpha \gamma^{\mu} P \nu_\beta)(\bar{f} \gamma_\mu P f),
\end{eqnarray}
Here $P$ belongs to the set $ \{P_{R},P_L\}$, where $P_R$ and $P_L$ represent the right and left-handed chirality operators, respectively. These operators are defined as $P_{R,L}=(1\mp \gamma^5)/2$. The subscripts $\alpha$ and $\beta$ correspond to different neutrino flavours, where $\beta$ can be $e$, $\mu$, or $\tau$, and $\{f\}$ represents a fermion, which can be $e$, $u$, or $d$.
The strength of NC-NSI is quantified by $\epsilon_{\alpha \beta }^{f}$, which serves as a dimensionless coefficient. This coefficient measures the interaction strength relative to the weak interaction coupling constant $G_{F}$.In other words, $\epsilon_{\alpha\beta}^{f, P}$ is typically of the order of $G_{x}/G_{F}$.

The Hamiltonian can be expressed as follows in the presence of NSI
\begin{eqnarray}\label{nsi2}
\mathcal{H}_{tot}=
\mathcal{H}_{vac}+\mathcal{H}_{mat}+\mathcal{H}_{NSI}~~~~~~~~~~~~~~~~~~~~~~~~~~~~~~~~~~~~\\=
\begin{pmatrix}
E_{1} & 0 & 0 \\ 
0 & E_{2}& 0 \\
0 & 0 & E_{3}
\end{pmatrix}+U^{\dagger } A\begin{pmatrix}
1+\epsilon_{ee}(x) &\epsilon_{e\mu}(x)  &\epsilon_{e\tau}(x) \\ 
\epsilon_{\mu e}(x) & \epsilon_{\mu \mu}(x) & \epsilon_{\mu \tau}(x)\\
 \epsilon_{\tau e}(x) & \epsilon_{\tau \mu}(x) & \epsilon_{\tau \tau}(x)
\end{pmatrix} U\notag.
\end{eqnarray}

The NSI parameters $\epsilon_{\alpha\beta}(x)$, where $\alpha,\beta=e, \mu, \tau$, represent non-standard interactions. They can be expressed as
\begin{equation}\label{nsi3}
    \epsilon _{\alpha \beta }(x)=\sum_{f=e,u,d}\frac{N_{f} (x) }{N_{e}(x)} \epsilon _{\alpha \beta }^{f}.
\end{equation}
Here, $x$ denotes the distance traveled by the neutrinos, and $N_{f}(x)$ represents the matter fermion density. Adhering to the charge neutrality condition ($N_p=N_e$) and drawing from the quark structure of neutrons and protons, relationships can be derived as $N_{u}(x) = 2 N_{p}(x) + N_{n}(x)$ and $N_{d}(x) = N_{p}(x) + 2 N_{n}(x)$. Substituting these conditions into Equation \eqref{nsi3}, the expression for $\epsilon_{\alpha\beta}(x)$ turns out to be
\begin{equation}
    \epsilon_{\alpha \beta }(x)=\epsilon_{\alpha \beta }^{e}+(2+Y_{n}(x))\epsilon_{\alpha \beta }^{u}+(1+ 2 Y_{n}(x))\epsilon_{\alpha \beta }^{d},
\end{equation}
where $Y_{n}=N_n(x)/N_e(x)$. NSI parameters can be complex as well as real. For complex NSI parameters, the flavor non-diagonal elements are not equal. For real NSI, $\epsilon_{\alpha\beta}=\epsilon_{\beta\alpha}$. NSI interactions can further be categorized into axial vector ($A$) and vector ($V$) types, where $\epsilon_{\alpha\beta}^f=\epsilon_{\alpha\beta}^{f,L} \pm \epsilon_{\alpha\beta}^{f,R}$ ('$-$' for axial vector, '$+$' for vector). Bounds on NSI parameters are derived from extensive global analyses of data from both oscillation and non-oscillation experiments \cite{Esteban:2018ppq, Esteban:2019lfo, Coloma:2019mbs, Coloma2023, Adriano}. \\

\subsection{Scalar NSI}
The neutrinos' interaction with matter is facilitated by the intermediary particles $W^{\pm}$ and $Z^{0}$, which significantly influence neutrino oscillations. These interactions manifest as an augmented potential term in the neutrino oscillation Hamiltonian. Additionally, the potential for scalar interactions of neutrinos presents an intriguing possibility. Neutrinos have the potential to engage in scalar interactions, coupling with a scalar, such as the Higgs boson, possessing a non-zero vacuum expectation value to acquire mass. This unconventional coupling between neutrinos and environmental fermions via a scalar is succinctly captured by the following lagrangian formulation \cite{parke, xu, khan, moon,denton, medhi,gupta, essnub}
\begin{equation}
\mathcal{L_{SNSI}}=y_{f}Y_{\alpha \beta }\left [  \bar{\nu _{\alpha }}(p_{3})\nu _{\beta }(p_{2}) \right ]\left [ \bar{f}(p_{1})f(p_{4}) \right ]
\end{equation}
where, the symbols $y_{f}$ and $Y_{\alpha\beta}$ denote the Yukawa couplings between the scalar mediator and the fermions  (represented by the subscript 'f') and the neutrinos, respectively. The modification introduced in the Lagrangian density, denoted by $\mathcal{L}$, induces a correction in the Dirac equation, manifested as follows
\begin{equation}
    \bar{\nu_{\beta } }\left [ \iota\partial _{\mu}\gamma _{\mu}+\left(M_{\beta \alpha }+\frac{\sum_{f}n_{f}y_{f}Y_{\alpha \beta }}{m_{\phi }^{2}}\right)  \right ]\nu _{\alpha }=0
\end{equation}
where, the mass of the scalar mediator, denoted as $m_{\phi}$, serves as a key parameter in this context. scalar non-standard interactions (SNSI) can manifest as adjustments to the neutrino mass matrix. Consequently, the effective Hamiltonian, when considering the influence of Scalar NSI, can be formulated as
\begin{equation}
    \mathcal{H}\approx E_{\nu }+\frac{M_{eff}M_{eff}^{\dagger }}{2E_{\nu }}\pm V_{m}
\end{equation}
where, $V_{m}$ is the matter potential term and $M_{eff}=M+M_{SNSI}$ with $M_{SNSI}\equiv \sum_{f}n_{f}y_{f}Y_{\alpha \beta }/m_{\phi }^{2}$.
The $\nu$-mass matrix can be diagonalized by a modified
mixing matrix as ${U}'=PUQ^{\dagger }$, where the Majorana rephasing matrix Q can be absorbed by $QD_{\nu}Q^{\dagger }=D_{\nu }=diag\left ( m_{1}, m_{2}, m_{3} \right )$. The unphysical diagonal rephasing matrix, P can be rotated away into the SNSI contribution
as follows,
\begin{equation}
    M_{eff}=UD_{\nu }U^{\dagger }+P^{\dagger }M_{SNSI}P=M+\delta M
\end{equation}
We parametrize the SNSI contribution $\delta M$ in a model-independent way as follows,
\begin{equation}\label{spara}
    \delta M=\sqrt{\left | \Delta m_{31}^{2} \right |}\begin{pmatrix}
\eta _{ee} & \eta _{e\mu} &\eta _{e\tau} \\ 
\eta _{e\mu}^{\ast } & \eta _{\mu\mu} & \eta _{\mu\tau}\\ 
\eta _{e\tau}^{\ast } & \eta _{\mu\tau}^{\ast } &\eta _{\tau\tau} 
\end{pmatrix},
\end{equation}
where the dimensionless parameter $\eta_{\alpha\beta}$ characterizes the magnitude of scalar non-standard interaction (SNSI), while $\sqrt{\left | \Delta m_{31}^{2}\right |}$ serves as a scaling factor. {\color{black} The parameters 
\(\eta_{\alpha\beta}\) in Eq. \ref{spara} are generally complex, 
\begin{equation}
\eta_{\alpha\beta} = |\eta_{\alpha\beta}| e^{i \phi_{\alpha\beta}},
\end{equation}
where \(\phi_{\alpha\beta}\) are the associated phases. In our numerical analysis, the complexity is explicitly taken into account. 
The effective Hamiltonian, and consequently the time-evolution operator and oscillation probabilities, depend on the full complex structure of the scalar NSI contribution \(\delta M\). For the results presented in this work, when evaluating the impact of a specific parameter \(\eta_{\alpha\beta}\), 
its magnitude is set to the $3\sigma$ upper bound from Table~\ref{Tab2}, and its phase \(\phi_{\alpha\beta}\) is set to the corresponding value provided in the same table. All other \(\eta\) parameters are set to zero in each individual case.} 

{\color{black}Equation~\ref{spara} can be traced back to a Yukawa-type interaction in which a scalar
mediator~$\phi$ couples to both background fermions~$f$ and neutrino flavors
through couplings~$y_f$ and~$Y_{\alpha\beta}$, respectively. Integrating out the
mediator induces an effective correction to the neutrino mass matrix of the form
$\delta m_{\alpha\beta} \simeq \sum_f n_f y_f Y_{\alpha\beta}/m_\phi^2$, where
$n_f$ is the fermion number density and $m_\phi$ is the mediator mass.

The dimensionless parameters~$\eta_{\alpha\beta}$ defined in Eq.~(31) represent
these effective corrections normalized to the oscillation mass scale. Hence,
$\eta_{\alpha\beta}$ provide a compact, model-independent way to describe the
impact of scalar NSI on neutrino oscillations. In principle, constraints on
$\eta_{\alpha\beta}$ from oscillation data can be translated into bounds on the
microscopic couplings~$y_fY_{\alpha\beta}/m_\phi^2$ once the mediator mass and
background composition are specified.
}

In the conventional scenario, a shared $m_{i}^{2}$ term permits the isolation of neutrino oscillations' reliance on mass-squared differences, namely $\Delta m_{21}^{2}$ and $\Delta m_{21}^{2}$. However, the introduction of SNSI introduces cross terms, such as $M\delta M^{\dagger}$ and $M^{\dagger}\delta M$, which preclude the subtraction of a common term from the mass matrix. Consequently, oscillation probabilities acquire a direct dependence on the absolute masses of neutrinos in the presence of SNSI. This study capitalizes on this direct mass dependence to establish constraints on the absolute neutrino mass.

\begin{table}[t]
	\centering
	\caption{Standard neutrino Oscillation parameters provided by NuFIT 5.2 (2022) (with SK atmospheric data) \cite{nufit} }
	\label{Tab1}
	\begin{tabular}{|c|c|c|c|}
		\hline
		Parameters  &  Best fit& Range (1$\sigma$)& Range (3$\sigma$)\\
		\hline\hline
		$\theta_{12}^{o}$    &   $33.4$& 32.66$\to$34.14 &  31.28$\to$35.75\\
		\hline
		$\theta_{13}^{o}$    & $8.58$  & 8.47$\to$8.68  & 8.23$\to$8.90 \\
		\hline
		$\theta_{23}^{o}$     &  $42.23$ & 41.14$\to$43.31  & 39.7$\to$50.99 \\
		\hline
		$\Delta m_{21}^{2}\times 10^{-5}\, \rm (eV^{2})$   & $7.41$  &7.20$\to$7.62   & 6.82$\to$8.03 \\
		\hline
		$\Delta m_{31}^{2}\times 10^{-3}\, \rm (eV^{2})$   & $2.507$  & 2.48$\to$2.53  & 2.427$\to$2.590 \\
		\hline
        $\delta_{CP}^{o}$   & $232.04$  & 195.95$\to$268.14  & 143.81$\to$350.07 \\
		\hline
	\end{tabular}
\end{table}

\begin{table}[t]
	\centering
	\caption{Scalar NSI parameters extracted from \cite{essnub} are given here, and these parameters are assumed to be complex.}
	\label{Tab2}
	\begin{tabular}{|c|c|c|}
		\hline
        NSI Parameters  & Range ($1 \sigma$)& Range ($3 \sigma$)\\      
		\hline\hline
				$\eta_{ee}$ &  -0.017$\to$0.018   & -0.036$\to$0.036 \\
         \hline
         $\eta_{\mu \mu}$ &-0.019$\to$0.012  & -0.051$\to$0.051 \\
		\hline
        $\eta_{\tau \tau}$ & -0.014$\to$0.015 &-0.039$\to$0.042\\
		\hline
		$\eta_{e \mu}$ & 0.000$\to$0.022  & 0.000$\to$0.135 \\
		\hline
		$\eta_{e \tau}$ & 0.000$\to$0.036 &  0.000$\to$0.196\\
		\hline
		$\eta_{\mu \tau}$ & 0.000$\to$0.024 &  0.000$\to$0.828 \\
		\hline
		$\phi_{e \mu}$ & -  & 108 Degree \\
		\hline
		$\phi_{e \tau}$ &- & -43.2 Degree\\
		\hline
		$\phi_{\mu \tau}$ & -& -158.4 Degree \\
		\hline
	\end{tabular}
\end{table}

To assess this, we systematically incorporate non-zero diagonal $\eta_{\alpha\beta}$ parameters individually. As an illustration, we present the effective mass matrix $M_{eff}$ under the condition $\eta_{ee}\neq 0$, highlighting its explicit reliance on the absolute neutrino masses.
\begin{equation}
   M_{eff}=Udiag\left(m_{1},m_{2},m_{3}\right)U^{\dagger }+\sqrt{\left | \Delta m_{31}^{2}\right |} diag\left ( \eta _{ee},0,0 \right ) 
\end{equation}

In the preceding discussion, we have elucidated quantum correlation measures both in terms of probabilities and through the lens of the reduced density matrix. It is apparent that these metrics can be readily computed using the formalism outlined in this section.



\section{Results and Discussion}\label{sec4}

In this section, we explore the impact of NSI on various quantum correlation measures within the context of the DUNE experimental setup. We focus on analyzing nonlocality measures, particularly the LGtI and CHSH inequalities. This analysis considers the influence of both vector and scalar NSI, with a special emphasis on scalar NSI. We evaluate the effect of each scalar NSI parameter on different quantum correlation measures individually.

{\color{black} We consider the DUNE (Deep Underground Neutrino Experiment) configuration for our numerical analysis. DUNE is an experimental facility which employs the NuMI neutrino beam with an energy range of 1--10~GeV from Fermilab and has a long baseline of 1300~km to the Sanford Underground Research Facility. This corresponds to $L/E \sim 10^3$~km/GeV, which ensures good sensitivity to leptonic CP violation and to the determination of the neutrino mass hierarchy \cite{dune}. The initial neutrino beam consists of muon neutrinos ($\alpha = \mu$), and the detectors primarily measure electron (anti)neutrino appearance ($P_{\mu e}$), with muon-neutrino disappearance ($P_{\mu\mu}$) also accessible in later phases.}

The values of the standard oscillation parameters are provided in table \ref{Tab1}. {\color{black}The vector NSI parameters used in our analysis are taken from the global fit presented in Ref. \cite{Esteban:2019lfo}. Specifically, we adopt the $3 \sigma$ upper bounds on the allowed ranges, extracted from Fig. 1 of Ref. \cite{Esteban:2019lfo}. This choice ensures that the parameter values remain within present experimental limits while allowing us to probe the largest potential impact of NSI effects on neutrino oscillation observables. The specific parameter values adopted in our calculations: $\epsilon_{ee}-\epsilon_{\mu \mu} = 1.7$,  $\epsilon_{\tau \tau}-\epsilon_{\mu \mu} = 0.9$, $|\epsilon_{e\mu}| = 0.25$, $|\epsilon_{e\tau}| = 0.6$, and $|\epsilon_{\mu\tau}| = 0.11$, $\phi_{e \mu} = 360^\circ$, $\phi_{e \mu} = 360^\circ$, $\phi_{\mu \tau} = 360^\circ$.}

The $1\sigma$ and $3\sigma$ ranges for the scalar NSI parameters are listed in table \ref{Tab2}. All analyses are conducted under the assumption of normal mass ordering, where the lightest neutrino mass eigenstate has a value of $7.42 \times 10^{-5}$ eV {\color{black}for consistency with the source of our scalar NSI parameters \cite{essnub}.}

\begin{figure*}[htb]
\includegraphics[scale=0.40]{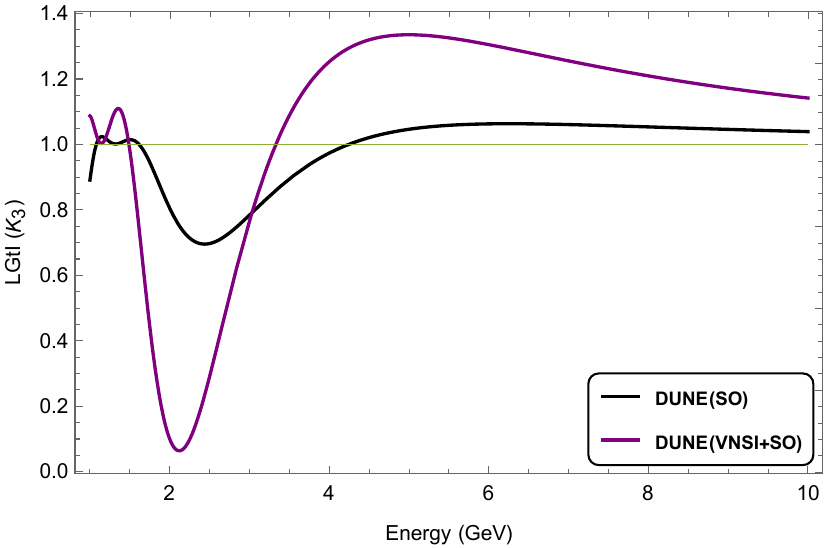}
\includegraphics[scale=0.40]{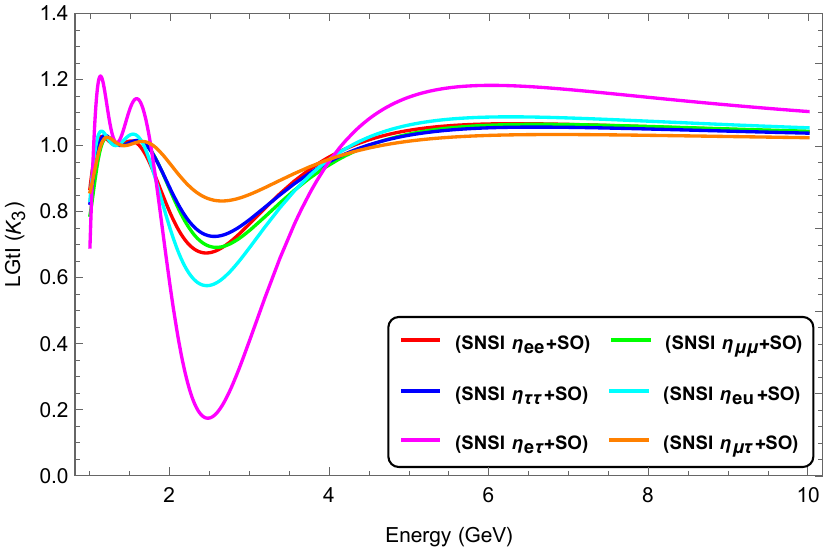}
\includegraphics[scale=0.40]{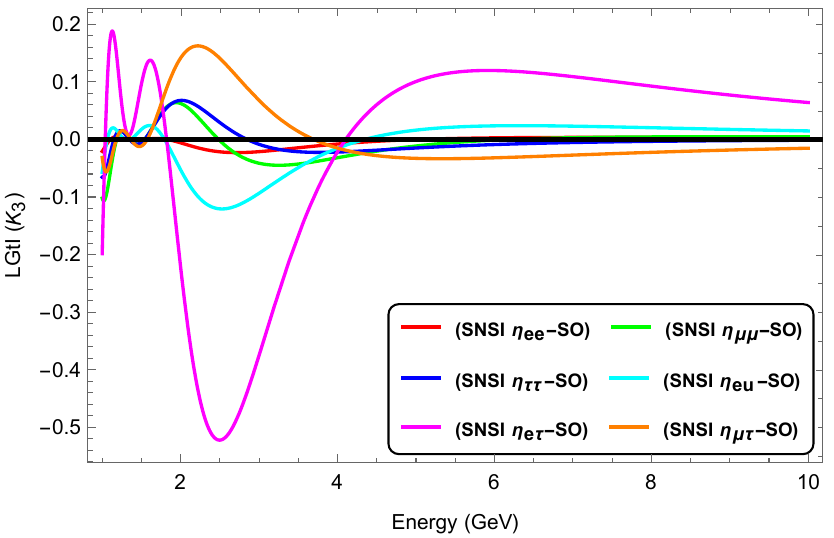}
\\
\includegraphics[scale=0.40]{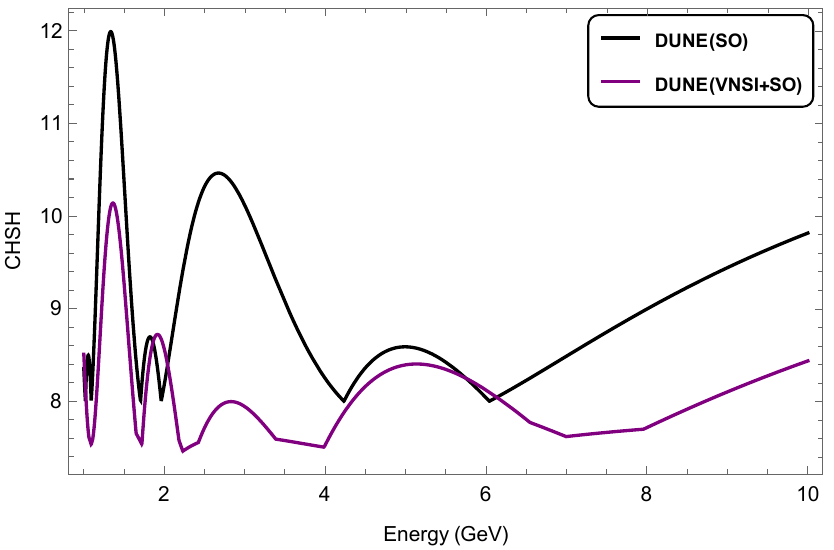}
\includegraphics[scale=0.40]{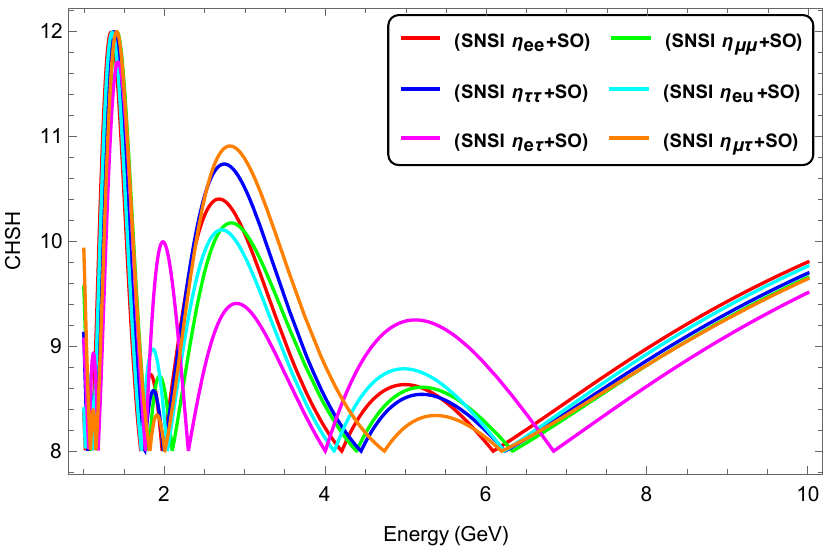}
\includegraphics[scale=0.40]{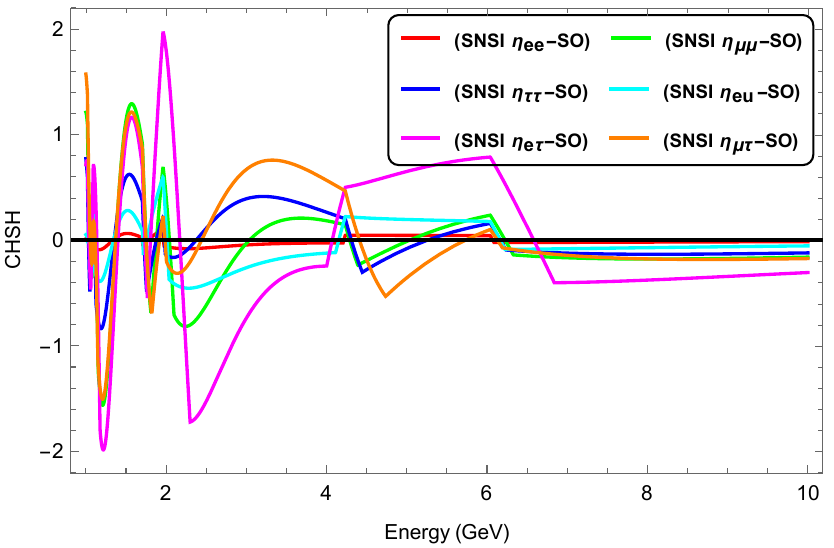}
\caption{Nonlocality measures, i.e., LGtI $(K_{3})$ (upper panel) and CHSH (trade-off relation) (lower panel), are plotted as functions of neutrino energy for the $SO$ and $VNSI+SO$ (left), $SNSI+SO$ (center), and $SNSI-SO$ (right) scenarios \cite{essnub}. These plots are generated using the $3\sigma$ upper bounds of the SNSI parameters within the DUNE experimental setup {\color{black}(baseline $L=1300$ km, beam energy range 1–10 GeV). The initial state is taken to be a muon neutrino ($\alpha=\mu$).} The standard neutrino oscillation parameters and NSI parameters are listed in Table~\ref{Tab1} and Table~\ref{Tab2}, respectively. {\color{black}The scalar NSI effects are shown for the $3 \sigma$ upper bounds from Table~\ref{Tab2}}. The color scheme is as follows: black for SM, purple for VNSI, red for SNSI ($\eta_{ee}$), green for SNSI ($\eta_{\mu \mu}$), blue for SNSI ($\eta_{\tau \tau}$), cyan for SNSI ($\eta_{e \mu}$), magenta for SNSI ($\eta_{e \tau}$), and orange for SNSI ($\eta_{\mu \tau}$).
}
\label{fig1}
\end{figure*}

Figure \ref{fig1} illustrates the behavior of non-locality quantum correlation measures for the three-flavor neutrino oscillation scenario. In the upper panel, we depict the temporal correlation measure of LGtI inequality, while in the lower panel, we showcase the spatial correlation measure of nonlocality, i.e., CHSH inequality. The left panel illustrates the nonlocality measures for the standard matter oscillation (SO) and the (VNSI+SO) scenario, highlighting the combined effects of matter interaction and vector non-standard interactions. The center panel presents the (SNSI+SO) scenario, capturing the influence of matter effects alongside scalar non-standard interactions. Finally, the right panel focuses solely on the effect of SNSI parameters, i.e., the (SNSI-SO) scenario.

The left panel highlights the discernible differences between the two scenarios, providing a clearer understanding of the impact of vector NSI on nonlocality measures. The black curve represents the SO scenario, while the purple curve corresponds to the (VNSI+SO) scenario. Consistently, LGtI is violated for energies exceeding 4 GeV in both the SO and (VNSI+SO) scenarios. The inclusion of vector NSI amplifies this violation of LGtI at higher energies. The violation of LGtI increases by approximately $20\%$ compared to the SO for the higher energy range. {\color{black} In the CHSH inequality, the inclusion of vector NSI decreases the value of the CHSH inequality across the entire energy range.}

In the central panel of figure \ref{fig1}, the graphical representation illustrates the impact of scalar NSI parameters on nonlocality measures. In this analysis, we evaluate the influence of each parameter on quantum correlation measures individually. The inclusion of the scalar $\eta_{e \tau}$ NSI parameter significantly enhances the violation of LGtI at higher energies. To clearly observe the effects of scalar NSI, the right panel exclusively depicts the (SNSI-SO) scenario, focusing solely on the influence of scalar NSI. Among the parameters, $\eta_{e \tau}$ is the most significant, leading to an approximate 10\% increase in LGtI violation compared to the SO scenario in the higher energy range. The $\eta_{e \mu}$ scenario also enhances LGtI violation at higher energies, though the effect is less pronounced than in the $\eta_{e \tau}$ scenario, while other parameters do not show LGtI violation. The $\eta_{e e}$ parameter is the least significant for LGtI. At E $\approx$ 2.5 GeV, which corresponds to the maximum neutrino flux at DUNE, the $\eta_{\mu \tau}$ scenario exhibits a more noticeable effect. {\color{black} The CHSH inequality is similarly enhanced by the $\eta_{e \tau}$ scenario. Additionally, near the energy corresponding to the maximum flux at DUNE, the $\eta_{\mu \tau}$ scenario shows a comparable enhancement of this inequality. At an energy of approximately 1.5 GeV, the CHSH inequality increases for all scalar NSI parameters.}

\begin{figure*}[htb]
\includegraphics[scale=0.40]{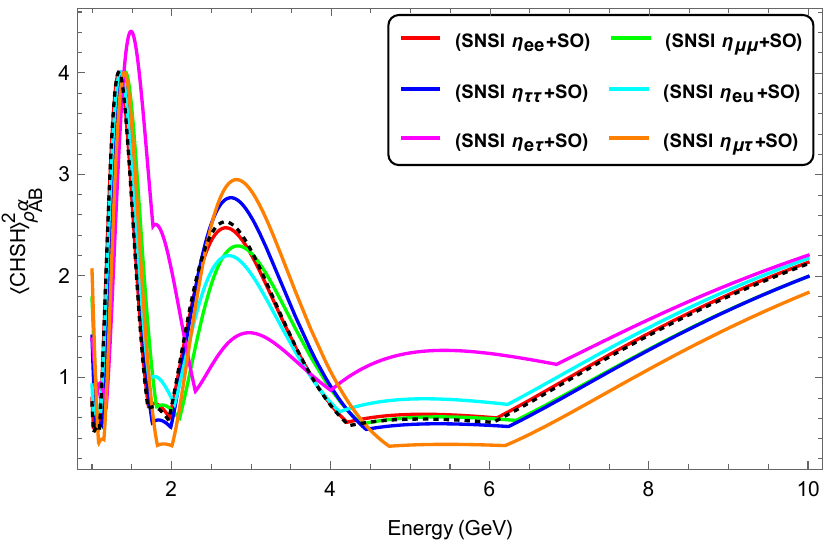}
\includegraphics[scale=0.40]{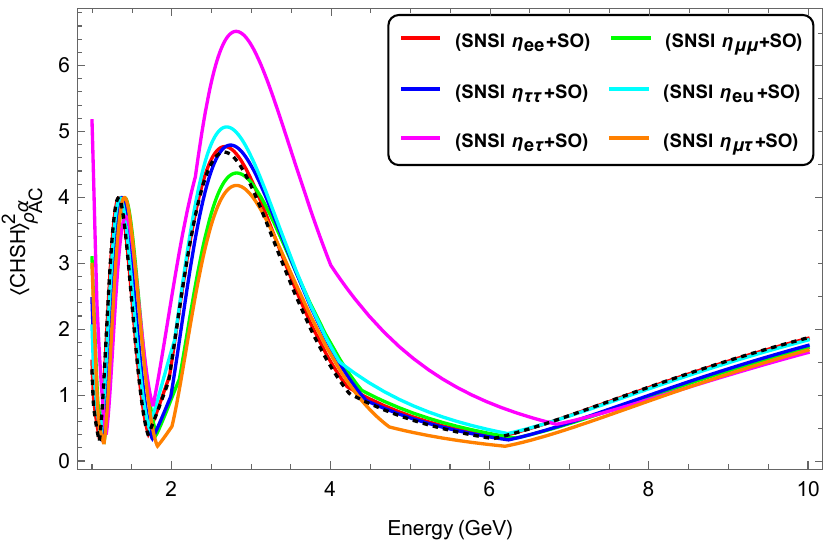}
\includegraphics[scale=0.40]{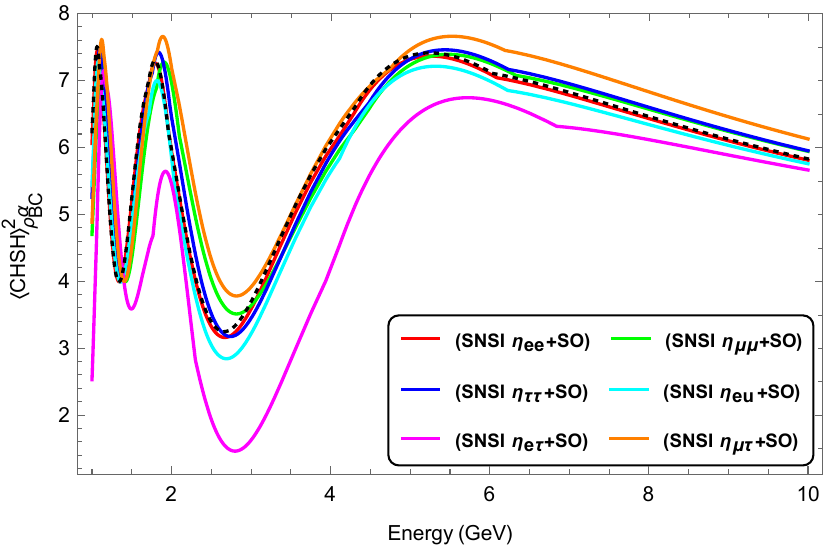}
\\
\includegraphics[scale=0.40]{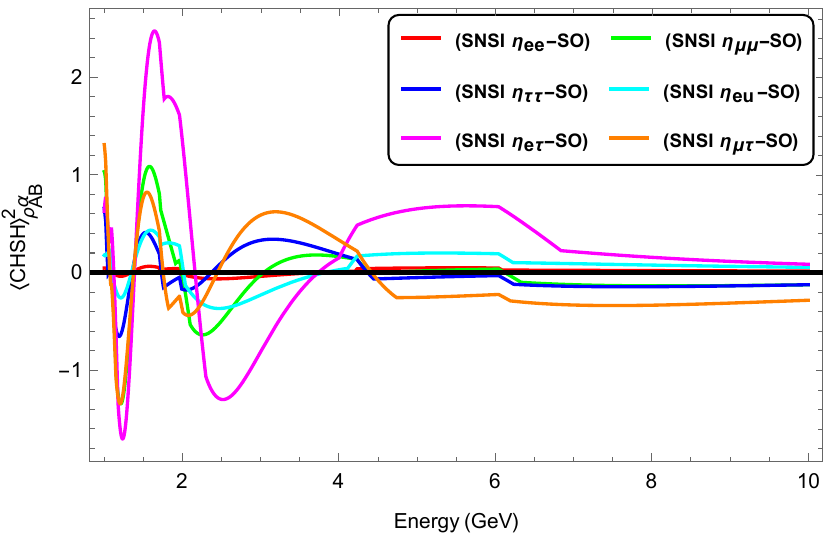}
\includegraphics[scale=0.40]{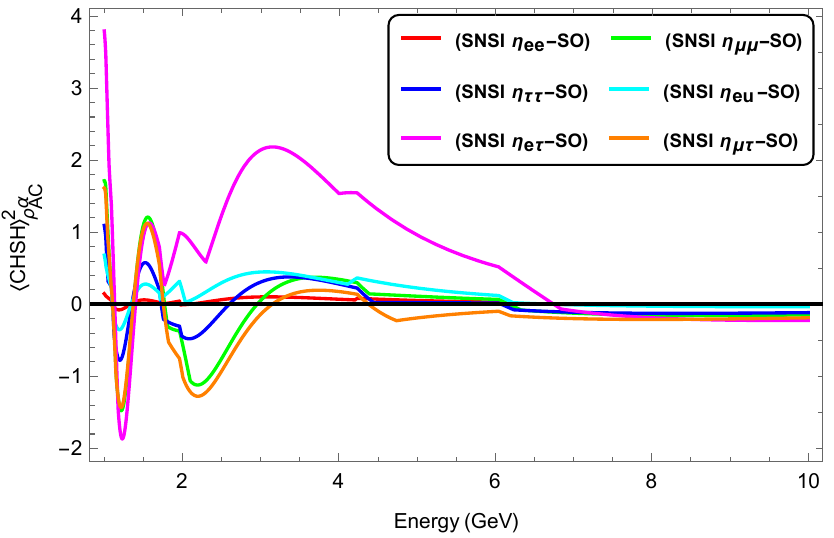}
\includegraphics[scale=0.40]{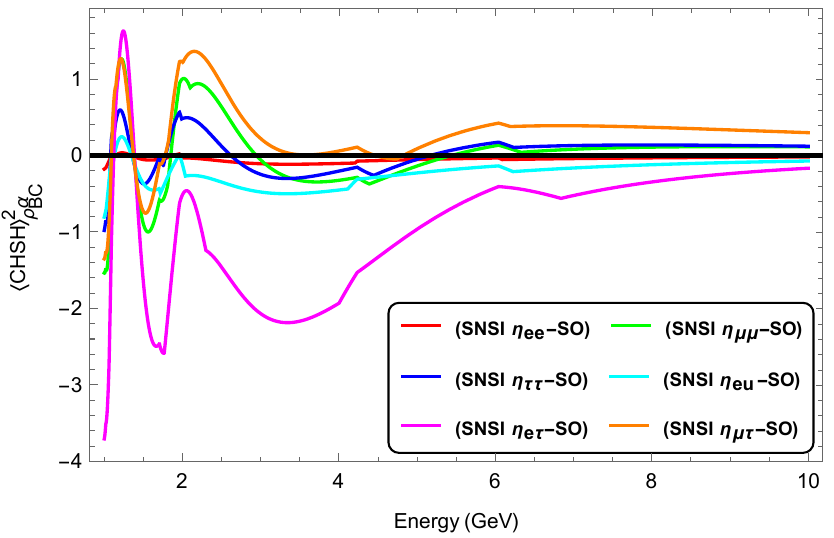}
\caption{ {\color{black}Spatial non-locality measures for the reduced two-qubit neutrino flavor states. The upper panels show the squares of the maximal CHSH expectation values, $\langle \text{CHSH} \rangle^2$, for the bipartite subsystems AB, AC, and BC, showing the combined effect of standard oscillations and scalar NSI. The lower panels show the corresponding pure scalar NSI effect, calculated as the difference from the standard oscillation case ($\text{SNSI} - \text{SO}$). All curves are for an initial muon neutrino beam ($\alpha = \mu$) and are plotted as a function of neutrino energy for the DUNE baseline. The scalar NSI parameters are set to their $3\sigma$ upper bounds from Table~\ref{Tab2}. In the upper panels, the black dotted curve represents the Standard Model case with matter effects (SO).} {\color{black} In the upper panels, values of $\langle{\rm CHSH}\rangle^{2}$ exceeding 4 indicate a violation of the Bell–CHSH inequality.}
}
\label{fig2}
\end{figure*}

\begin{figure*}[htb]
\includegraphics[scale=0.40]{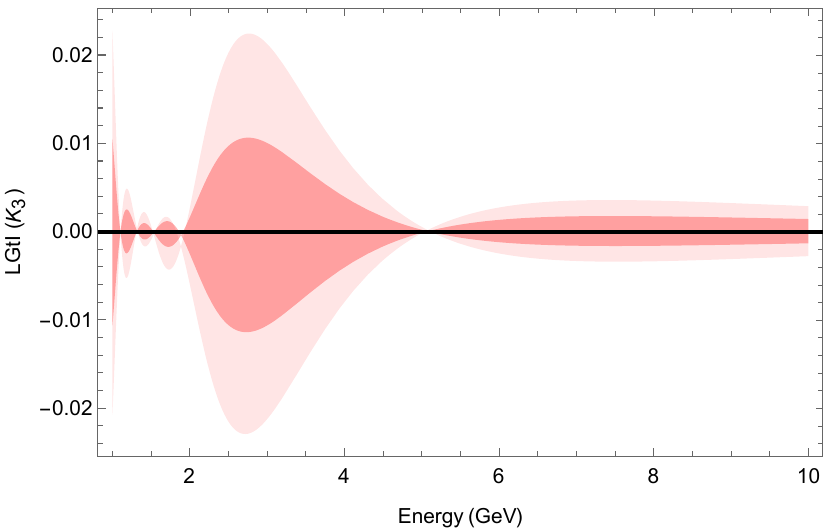}
\includegraphics[scale=0.40]{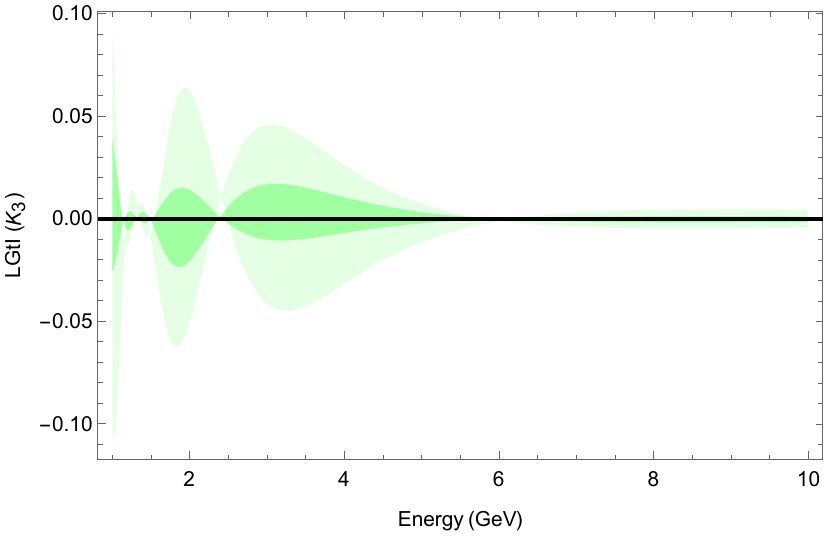}
\includegraphics[scale=0.40]{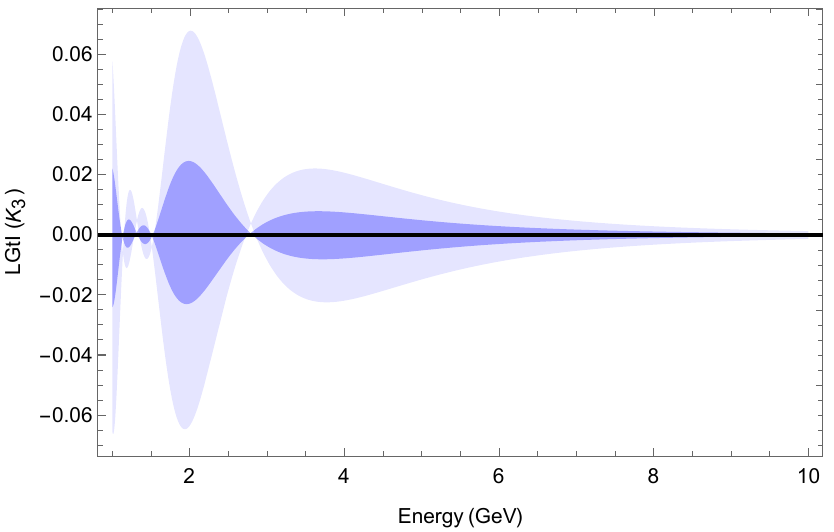}
\caption{The plots show LGtI $(K_{3})$ versus neutrino energy for SNSI parameters: $\eta_{ee}$ (left), $\eta_{\mu \mu}$ (center), and $\eta_{\tau \tau}$ (right) for the (SNSI-SO) scenario. The darker regions represent the $1\sigma$ range {\color{black}(upper bound and lower bound)}, and the lighter regions indicate the $3\sigma$ range for the SNSI parameters given in Table \ref{Tab2}. {\color{black}Other parameters are same as mentioned in Figure \ref{fig1}.}}
\label{fig4}
\end{figure*}

The bipartite spatial non-locality, quantified by the Bell-CHSH inequality, is presented in Figure \ref{fig2} for all three reduced two-qubit subsystems (AB, AC, BC). The upper panels show the combined effect of standard oscillations and scalar NSI (SNSI $+$ SO), while the lower panels isolate the pure NSI contribution (SNSI $-$ SO).

{\color{black}
However, the scalar NSI parameters do induce measurable modifications to the correlation strengths. The most significant effects are observed in the BC subsystem, where several parameters, particularly $\eta_{\mu\tau}$ and $\eta_{e\tau}$, cause substantial deviations from the standard oscillation case (black dotted curve). The lower panels confirm that these are genuine NSI effects, with $\eta_{e\tau}$ producing the most pronounced pure contribution across multiple subsystems.

Building on the general results from Figure~\ref{fig2}, a detailed analysis of the numerical values reveals distinct behaviors across the different bipartite subsystems. { \color{black}The maximum values of $\langle \text{CHSH} \rangle^2$ in the Standard Model (SO) case are approximately 4.0 for AB, 4.6 for AC, and 7.6 for BC, confirming that the BC subsystem harbors the strongest bipartite correlations.}

The introduction of scalar NSI parameters modifies these correlations in a subsystem-dependent manner. {\color{black}The parameter $\eta_{e\tau}$ provides a significant positive enhancement, increasing the maximum values in the AB and AC subsystems to approximately 2.5 and 3.8, respectively.} Conversely, in the BC subsystem, the effect of $\eta_{e\tau}$ is to reduce the correlation strength. In contrast, the parameter $\eta_{\mu\tau}$ exhibits a strong positive contribution specifically within the BC subsystem, further amplifying its already significant correlation value.

These observations confirm that scalar NSI does not merely provide a universal scaling of quantum correlations but instead couples selectively to specific flavor pairs, enhancing correlations in some subsystems while suppressing them in others.} {\color{black}Such enhancement in one subsystem accompanied by suppression in
another is consistent with the trade-off relation of bipartite non-locality in a tripartite system. Therefore, even when individual
subsystems display Bell-CHSH violation, the overall three-flavor neutrino
state retains its intrinsically tripartite quantum character.
}

Figure \ref{fig4} showcases the LGtI measure across a range of diagonal scalar NSI parameters, delineating $1 \sigma$ and $3 \sigma$ error bars. In this figure, we examine the (SNSI-SO) scenario. The darker region corresponds to the $1 \sigma$ error bar, while the lighter region corresponds to the $3 \sigma$ error bar of scalar NSI parameter. The color scheme is as follows: red represents $\eta_{ee}$, green represents $\eta_{\mu \mu}$, and blue represents $\eta_{\tau \tau}$. Here the enhancement of LGtI for $\eta_{\mu \mu}$ and $\eta_{\tau \tau}$ from 1 sigma to 3 sigma is approximately 0.04, while for $\eta_{e e}$, it is 0.01. Thus, the contributions of $\eta_{\mu \mu}$ and $\eta_{\tau \tau}$ are greater than that of $\eta_{ee}$. Here, we simply showcase the LGtI correlation measure as the scalar NSI parameters vary and the impact of the $\eta_{ee}$ parameter is the least pronounced.

{\color{black} Inspired by the potential to derive constraints from quantum violations, we systematically explored the parameter space of Scalar NSI. We found that while parameters like $\eta_{e\tau}$ significantly enhance the pre-existing LGtI violation, the fact that the inequality is already violated in the Standard Model scenario prevents us from establishing a new, stricter upper bound through this particular method. This intriguing approach, however, remains highly promising for probing other BSM scenarios where the classical bound is not violated in the SM limit.}

In this study, we utilize {\color{black}  scalar NSI parameters constrained by the ESSnuSB experiment \cite{essnub}} to investigate the impact of scalar NSI on quantum correlation measures, particularly LGtI and CHSH inequalities. This is the first time SNSI effects are being explored in the context of such quantum observables. Since SNSI parameters are still weakly constrained, the resulting correlations are sensitive to the choice of bounds, motivating a detailed analysis under different upper limits. Additionally, the high-precision sensitivity of the DUNE experiment to both standard oscillation parameters and potential NSI effects makes it a complementary setup for probing new physics signatures such as SNSI.  

{\color{black}Recent works have explored scalar NSI effects using oscillation probabilities \cite{Sarker1, Sarker2}. Their analyses show that the impact of $\eta_{\mu\mu}$ and $\eta_{\tau\tau}$ is stronger than that of $\eta_{ee}$, which is consistent with the trend we observe in our LGtI analysis (Figure \ref{fig4}). Our study extends these results by employing quantum correlation measures (LGtI and CHSH), which directly test the non-classicality of neutrino oscillations. Interestingly, we find that $\eta_{e\tau}$ leads to the most significant enhancement of non-locality violations, a feature not captured in conventional probability-based studies. This highlights the complementary nature of our approach.}

{\color{black}Therefore, quantum correlations serve as a powerful complement to standard neutrino oscillation probabilities, offering a distinct lens to scrutinize new physics, as demonstrated in our prior work \cite{Konwar:2024nwc}. We posit that this combined approach will be crucial for deeper understanding of the scalar NSI.}


\section{Conclusion}\label{sec5}
In this work, we investigated the effects of scalar NSI on both spatial and temporal measures of non-locality, with a primary focus on the CHSH and Leggett-Garg type inequalities. The analysis was conducted using the DUNE experimental setup. We assessed the influence of each scalar NSI parameter on quantum correlation measures individually. Our findings indicate that both vector and scalar NSI significantly enhance LGtI violations at higher energies, with the scalar parameter $\eta_{e \tau}$ emerging as the most dominant. While the $\eta_{e \mu}$ parameter also contributes to LGtI violations at higher energies, its effect is considerably weaker compared to $\eta_{e \tau}$, and other NSI parameters show no violations. The $\eta_{e e}$ parameter is the least influential for LGtI. At an energy around 2.5 GeV, which aligns with the peak neutrino flux at DUNE, the $\eta_{\mu \tau}$ scenario exhibits a more pronounced effect. {\color{black}Furthermore, the CHSH inequality is similarly enhanced in the $\eta_{e \tau}$ scenario, highlighting its importance. Furthermore, our analysis of the bipartite subsystems reveals that scalar NSI parameters, particularly $\eta_{e\tau}$ and $\eta_{\mu\tau}$, couple selectively to specific flavor pairs, enhancing spatial correlations in some subsystems while suppressing them in others.}

\bibliographystyle{apsrev4-2}

\title{References}

 \end{document}